\documentclass[twocolumn,preprintnumbers,superscriptaddress,endnote,nofootinbib,aps,prd,floatfix]{revtex4}

\usepackage[utf8]{inputenc}

\usepackage{footmisc,enumerate}
\usepackage{multirow}
\usepackage{subfigure}
\usepackage{amsmath,amssymb}
\usepackage{graphicx}
\usepackage{placeins}
\usepackage{bbm}
\usepackage{xspace,slashed}
\usepackage{hyperref}
\hypersetup{colorlinks=true, citecolor=blue, urlcolor=blue, linkcolor=blue}
\usepackage[normalem]{ulem}

\usepackage[autostyle]{csquotes}

\begin{document}

\newcommand{\og}{\ensuremath{\tilde{O}_g}\xspace}
\newcommand{\ot}{\ensuremath{\tilde{O}_t}\xspace}

\title{Probing the Symmetric Higgs Portal with Di-Higgs Boson Production}

\begin{abstract}
A coupling of a scalar, charged under an unbroken global U(1) symmetry, to the Standard Model via the Higgs portal is one of the simplest gateways to a dark sector. Yet, for masses $m_{S}\geq m_{H}/2$ there are few probes of such an interaction. In this note we evaluate the sensitivity to the Higgs portal coupling of di-Higgs boson production at the LHC as well as at a future high energy hadron collider, FCC-hh, taking into account the full momentum dependence of the process. This significantly impacts the sensitivity compared to estimates of changes in the Higgs-coupling based on the effective potential. We also compare our findings to precision single Higgs boson probes such as the cross section for vector boson associated Higgs production at a future lepton collider, e.g. FCC-ee, as well as searches for missing energy based signatures.
\end{abstract}

\author{Christoph Englert} \email{christoph.englert@glasgow.ac.uk}
\affiliation{SUPA, School of Physics \& Astronomy, University of Glasgow, Glasgow G12 8QQ, UK\\[0.1cm]}
\author{Joerg Jaeckel} \email{jjaeckel@thphys.uni-heidelberg.de}
\affiliation{Institut f\"ur Theoretische Physik, Universit\"at Heidelberg,
Philosophenweg 16, 69120 Heidelberg, Germany\\[0.1cm]}


\maketitle

\section{Introduction}
\label{sec:intro}
Searches for a dark sector are often benchmarked by the use of three portals coupling particles neutral under the Standard Model (SM) gauge group to SM particles: the vector portal, the neutrino portal and the Higgs portal~\cite{Binoth:1996au,Schabinger:2005ei,Patt:2006fw,Ahlers:2008qc,Batell:2009yf}.
On closer inspection most investigations (see e.g. the recent Ref.~\cite{Beacham:2019nyx}) actually focus on the respective mixing effects, i.e. kinetic mixing between a new vector and the SM U(1), neutrino mixing with a new neutral lepton and mixing of the Higgs boson with a dark counterpart.
However, in the case of the Higgs portal there is the possibility to have an unbroken U(1) or ${\mathbb{Z}}_{2}$ symmetry that forbids mixing effects.
Although there are notable exceptions (cf., e.g.~\cite{Curtin:2014jma,Craig:2014lda,Arcadi:2019lka} for overviews)  this case is much less explored and constrained. This is despite the fact that such a dark scalar could be a dark matter candidate both from thermal~\cite{Silveira:1985rk,McDonald:2001vt,Davoudiasl:2004be,Barger:2007im,Khoze:2014xha,Feng:2014vea,Athron:2018ipf,Arcadi:2019lka} and from non-thermal~\cite{Hardy:2018bph,Bernal:2018kcw,Alonso-Alvarez:2019pfe} production and could also play a role in baryogenesis~\cite{Espinosa:2007qk,Noble:2007kk,Espinosa:2008kw,Espinosa:2011ax,Curtin:2014jma}.

Let us consider the following simple Lagrangian for a dark scalar with an unbroken ${\mathbb{Z}}_{2}$ symmetry,
\begin{equation}
\label{eq:model}
{\mathcal{L}}=\frac{1}{2}(\partial_{\mu} S)^2-\frac{m^{2}_{S}}{2}S^2-\lambda S^2(\Phi^{\dagger}\Phi-v^2/2)\,,
\end{equation}
where $\lambda$ specifies the Higgs portal coupling with the SM Higgs doublet $\Phi$. 

Experimental searches without the dark matter assumption are complicated by the fact that such a dark scalar can only be pair produced via the exchange of at least one Higgs. In particular for dark scalar mass above the threshold, $m_{H}/2$, where the Higgs boson cannot decay into two scalars sensitivity is
severely limited.

In this note, we evaluate the impact of the virtual effects of such a dark scalar on Higgs boson pair production. While the LHC has relatively limited sensitivity, a future 100 TeV proton collider such as the so-called future circular hadron--hadron collider (FCC-hh) can viably test the unexplored parameter space for masses $m_{S}\leq m_{H}/2\leq {\rm few}\times 100\,{\rm GeV}$ given the enhanced search potential of 
Higgs pair final states, e.g.~\cite{Contino:2016spe}.
Compared to earlier studies~\cite{Noble:2007kk,Curtin:2014jma} we take into account the full, momentum-dependent one-loop amplitude (similar to what is done in~\cite{Voigt:2017vfz} for a different dark matter model interacting with the Higgs boson as well as~\cite{He:2016sqr} which considers a Higgs portal model in the non-mixing limit but one without a $\mathbb{Z}_{2}$ symmetry) instead of considering the correction of the Higgs boson self-coupling obtained from the effective Coleman-Weinberg potential~\cite{Coleman:1973jx}. In particular at low masses the threshold effects have significant impact, unfortunately somewhat reducing the actual sensitivity for some masses.

\section{Light scalar contribution to Higgs pair production}
\label{sec:calc}

\begin{figure}[!b]
\includegraphics[width=0.4\textwidth]{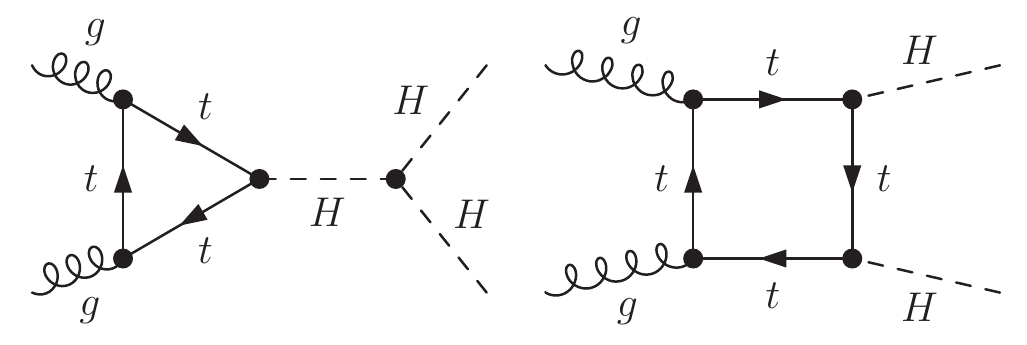}
\caption{Representative leading order Feynman topologies contributing to $gg \to HH$ production. \label{fig:lotops}}
\end{figure}

\begin{figure*}[!t]
\includegraphics[width=0.7\textwidth]{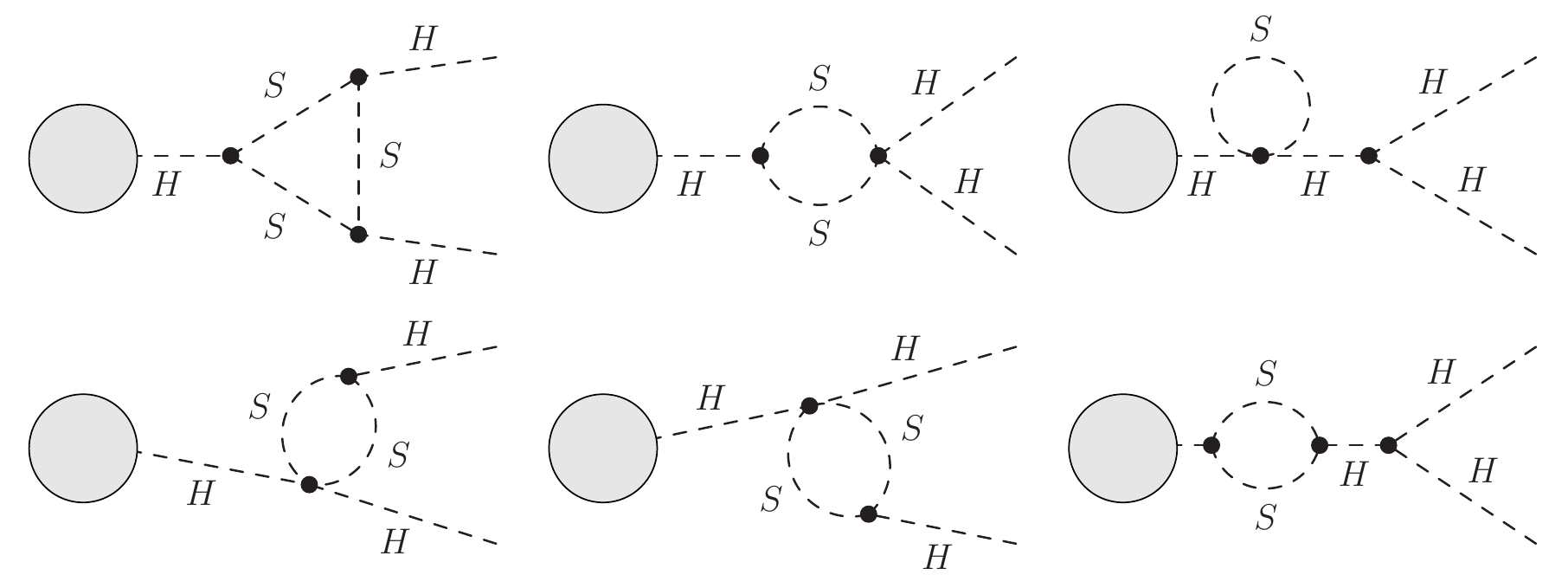}
\caption{One-loop contribution to $s$-channel $gg \to HH$ production. The shaded area represents the remainder (one-loop top insertion part) of the amplitude. \label{fig:nlotopss}}
\end{figure*}

\begin{figure*}[!t]
\includegraphics[width=0.7\textwidth]{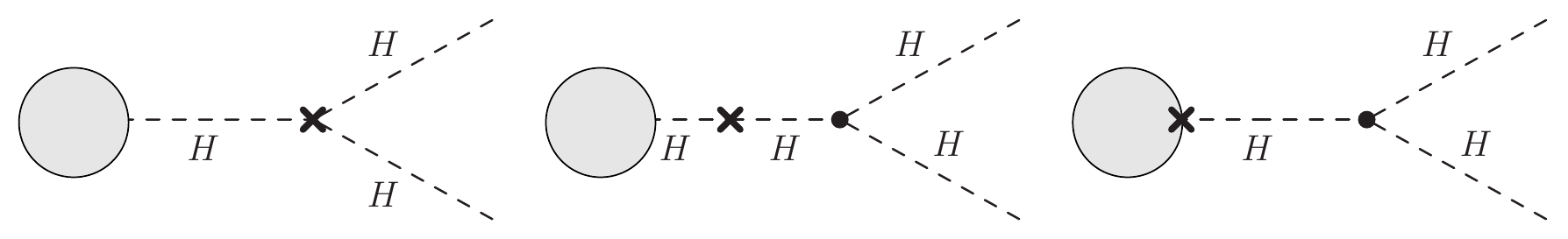}
\caption{Counter term contribution to $s$-channel $gg \to HH$ production. The shaded area represents the remainder of the amplitude as in Fig.~\ref{fig:nlotopss}. \label{fig:cttopss1}}
\end{figure*}

\begin{figure}[!b]
\includegraphics[width=0.25\textwidth]{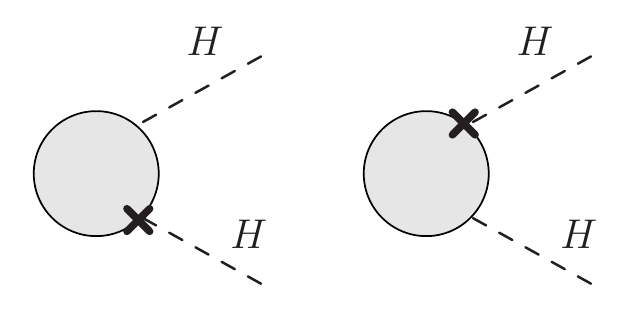}
\caption{Counter term contribution to the box graphs of $gg\to HH$ production. The shaded area represents the remainder (one-loop top insertion part) of the box amplitude. \label{fig:cttopss}}
\end{figure}
Let us start by explicitly writing down the portal interaction in terms of the Higgs, $H$, and the Goldstone modes $\phi^{\pm},\phi^0$.
Writing the Higgs doublet as
\begin{equation}
\label{eq:hdoub}
\Phi={1\over \sqrt{2}}\left(\begin{matrix} \sqrt{2} \phi_+ \\ {v+H+i\phi_0} \end{matrix}\right)\,,
\end{equation}
we have
\begin{multline}
\label{eq:new}
-{\cal{L}}_{\text{portal}}=  \lambda \Phi^\dagger \Phi S^2 
= +{\lambda v^2 \over 2} S^2 + {\lambda v } H S^2  + {\lambda \over 2} H^2S^2  \\
+ \frac{\lambda}{2} \phi_0^2 S^2 + \lambda \phi_+\phi_- S^2.
\end{multline}

Note that with the above definitions, we denote the $S$ particle's pole mass with $m_S$. The leading order contributions to $gg\to HH$ 
are given by the Feynman topologies shown in Fig.~\ref{fig:lotops}. At the same time, relevant one-loop $S$ contributions to the gluon fusion amplitude (modulo field renormalisation constants) are due to the off-shell Higgs three point function shown in Fig.~\ref{fig:nlotopss}. 

Let ${\cal{S}}$ be the $s$-channel and ${\cal{B}}$ the box part of the one-loop $gg\to HH$ amplitude, i.e. the left- and right-hand side of Fig.~\ref{fig:lotops} where all possible fermion flow orientations are understood implicitly. The full $gg\to HH$ matrix-element is then represented by
\begin{equation}
{\cal{M}} =  {\cal{S}} + {\cal{B}}.
\end{equation}
In the following we will consider the one-loop $S$ insertion for ${\cal{S}}$. Writing 
\begin{equation}
{\cal{S}} =  {{\cal{T}}} {1\over  s-m_H^2 } \Gamma(s,m_H^2,m_H^2)
\end{equation}   
where ${\cal{T}}$ denotes the well-known expression of one-loop Higgs boson production~\cite{Georgi:1977gs,Dawson:2018dcd}, with $s=m^2_{HH}=(p_{H,1}+p_{H,2})^2$, we can directly identify the leading order (or Born-level) contribution
\begin{equation}
\Gamma^{\text{Born}}(s,m_H^2,m_H^2)= -{3  m_H^2\over v} = -6\,\lambda^{\text{SM}}_H
\end{equation}
as the Higgs trilinear vertex in the SM. The virtual corrections induced by $S$ arise from the diagrams depicted in Fig.~\ref{fig:nlotopss} and are found to be
\begin{multline}
\label{eq:virt}
\Gamma^{\text{virt}} (s,m_H^2,m_H^2)=\\  {\lambda \over 16\pi^2 v} \bigg(
{3 m_H^2 A_0(m_S^2)  \over s-m_H^2} 
+ 2\lambda v^2 \bigg\{ 2 B_0(m_H^2,m_S^2,m_S^2) \\ + 4 \lambda v^2 C_0(m_H^2,m_H^2,s,m_S^2,m_S^2,m_S^2) \\
+ B_0 (s,m_S^2,m_S^2) \left[ 1+ {3m_H^2 \over s-m_H^2} \right] \bigg\} \bigg)\,.
\end{multline}
This contains divergences that are renormalised by the counter term contributions shown in Fig.~\ref{fig:cttopss1}. 
In Eq.~\eqref{eq:virt} $A_0,B_0,$ and $C_0$ are the well-known one-loop Passarino-Veltman~\cite{Passarino:1978jh} functions in the convention of Refs.~\cite{Denner:1991kt,Hahn:1998yk,Hahn:2000jm,Hahn:2000kx}.

Tadpoles deserve a special comment as they generate a non-vanishing contribution for the Higgs boson self-interaction renormalisation (see e.g.~\cite{Denner:1991kt}). The SM Higgs potential reads, after inserting Eq.~\eqref{eq:hdoub},
\begin{multline}
\label{eq:tadpole}
V_{\text{SM}} (\Phi) = -\mu^2 |\Phi|^2 + \lambda^{\text{SM}}_H  |\Phi |^4 \\ \supset v(-\mu^2 +v^2\lambda^{\text{SM}}_H)H  = t H\,.
\end{multline}
$t=v(-\mu^2 +v^2\lambda_H)$ vanishes at leading order due to the choice of $v$. Keeping track of $t=v(-\mu^2 +v^2\lambda_H)$
gives rise to a trilinear contribution
\begin{equation} 
\label{eq:extra}
V_{\text{SM}} (\Phi) \supset \left( {m_H^2\over 2 v} - {t\over 2v^2} \right) H^3 \,. 
\end{equation}
Our $S$-induced tadpole contributions can be removed through a choice that can be diagrammatically expressed as
\begin{equation}
\delta t + t = \delta t + \parbox{2.1cm}{\vspace{-0.1cm}\includegraphics[width=2.1cm]{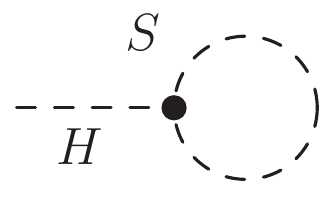}}= 0\,,
\end{equation}
which amounts to performing the calculation with the ``correct choice'' of $v$. This allows us to neglect tadpole diagrams as they would identically cancel against the associated counter terms. 
$\delta t$ is straightforward to compute in our scenario
\begin{equation}
\delta t = - {\lambda v\over 16\pi^2} \text{Re}\,A_0(m_S^2)  \,
\end{equation}
and needs to be considered in the renormalisation of the 3-point vertex function according to Eq.~\eqref{eq:extra}.
The counter term contribution is then given by
\begin{equation}
\Gamma^{\text{CT}} (s,m_H^2,m_H^2) = - {3\over v} \left({\delta t \over v} -  {\delta m_H^2 \,s\over s-m_H^2}  - {\delta Z_H m_H^2\over 2}  \right)
\end{equation}
where $\delta Z_H$ and $\delta m_H^2$ are the Higgs wave function and mass renormalisation constants.
We use the on-shell renormalisation scheme where these parameters are given by
\begin{equation}
\delta Z_H = -{\lambda^2 v^2 \over 8 \pi^2 } {\partial \over \partial q^2} \text{Re}\, B_0(q^2,m_S^2,m_S^2) \bigg|_{q^2=m_H^2}
\end{equation}
and 
\begin{equation}
\delta m_H^2 = {\lambda  \over 16 \pi^2 } \text{Re}\, A_0(m_S^2) + {\lambda^2 v^2\over 8\pi^2} \text{Re}\, B_0(m_H^2,m_S^2,m_S^2)\,,
\end{equation}
respectively. In dimensional regularisation with dimension $D<4$ 
we can extract the divergent contributions for $D\to 4$ via
\begin{equation}
A_0= {2m^2\over 4-D} + {\cal{O}}( [4-D]^0)\,,~B_0 = {2\over 4-D} + {\cal{O}}([4-D]^0)
\end{equation}
to realise that
\begin{equation}
\Gamma^{\text{virt}} + \Gamma^{\text{CT}} = {\cal{O}}( [4-D]^0)
\end{equation}
is manifestly UV finite. $\delta Z_H$ contains
no UV singularity. The full renormalised $s$-channel amplitude is then (see also \cite{He:2016sqr})

\begin{equation}
{\cal{S}}_{\text{virt}} =  {{\cal{T}} \over s-m_H^2} \left(\Gamma^\text{virt} + \Gamma^\text{CT} + {\delta Z_H \over 2}
\Gamma^{\text{Born}}\right)
\end{equation}
where the last term stems from the counter term contribution to the Higgs coupling as part of ${\cal{T}}$. The box contribution follows
analogously, Fig.~\ref{fig:cttopss}
\begin{equation}
{\cal{B}}_{\text{virt}} = \delta Z_H {\cal{B}}_{\text{Born}}\,.
\end{equation}
We obtain the full amplitude by expanding to ${\cal{O}}(\lambda^2)$
\begin{equation}
|{\cal{A}}|^2 = |{\cal{M}}_{\text{Born}}|^2 + 2\text{Re}\left( {\cal{M}}_{\text{Born}} {\cal{M}}^\ast_{\text{virt}}\right)\,.
\end{equation} 
As ${\cal{M}}_{\text{virt}}$ is UV-finite we can include the ${\cal{O}}(\lambda^4)$ term $|{\cal{M}}_{\text{virt}}|^2$ for comparison
to gauge the importance of (factorisable) two-loop contributions, in particular when we consider numerically large couplings $\lambda$ in our scan.

Through the choice of model, we implicitly assume that the dominant electroweak corrections indeed arise through the dynamics of $S$ and we will neglect the SM electroweak corrections, which are currently unknown, throughout. However, in direct relation to the SM expectation, i.e. when ratios are considered, the SM corrections will cancel at one-loop order. We have implemented the above calculation in a modified version of {\sc{Vbfnlo}}~\cite{Arnold:2008rz,Arnold:2011wj,Arnold:2012xn,Baglio:2014uba} that links the {\sc{FeynArts}}/{\sc{FormCalc}}/{\sc{LoopTools}}~\cite{Hahn:1998yk,Hahn:2000jm,Hahn:2000kx} to obtain numerical results.

\section{Results}
\begin{figure*}[!t]
\centering
\subfigure[~]{\includegraphics[width=0.45\textwidth]{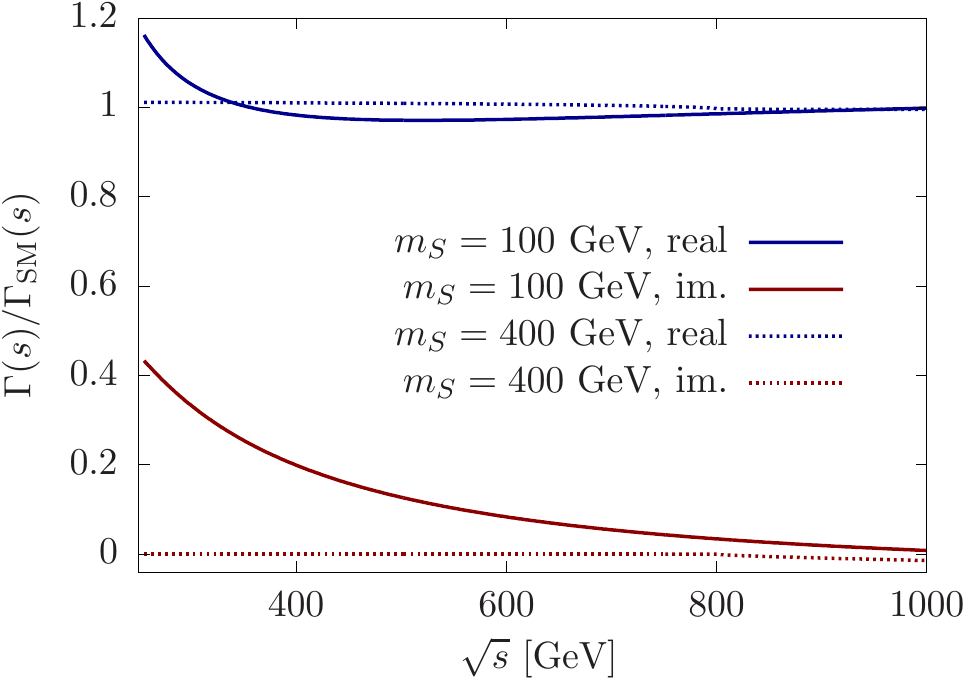}}
\hspace{1cm}
\subfigure[~]{\includegraphics[width=0.45\textwidth]{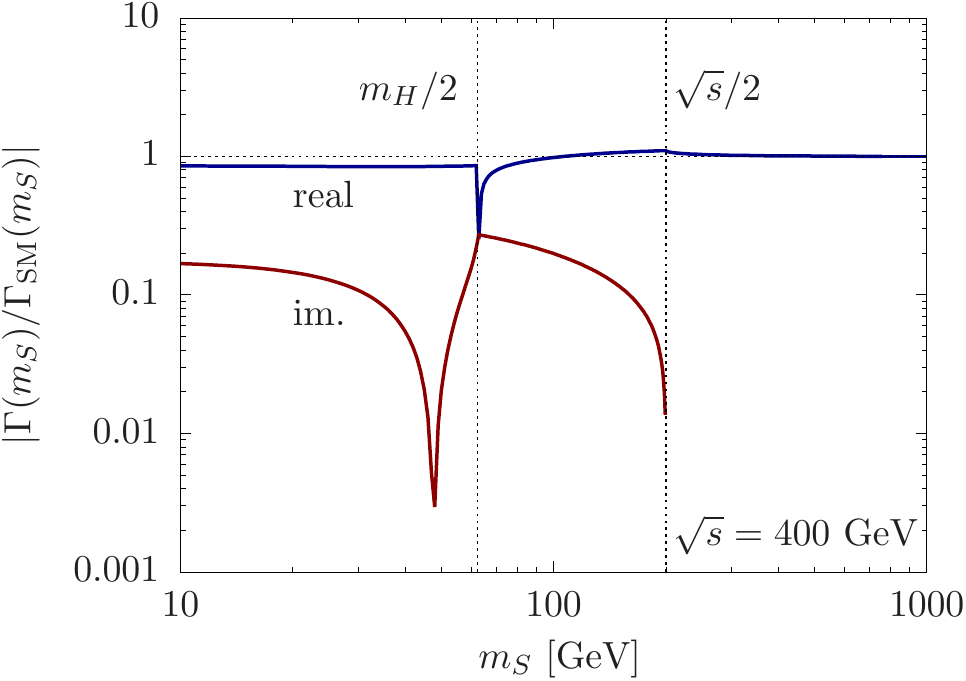}}
\caption{(a) Comparison of the imaginary and the real part of the three point function $\Gamma$ for ($\lambda=1$) relative to SM ($\lambda=0$) as a function of the invariant
di-Higgs mass $\sqrt{s}=m_{HH}$. The turn-on of absorptive parts is visible for $m_S=400$ GeV at $\sqrt{s}=m_{HH}=2m_S=800$~GeV.
(b) The {\emph{modulus}} of the three point function relative to the SM as a function of $m_S$ (again for $\lambda=1$) for fixed $\sqrt{s}=400$~GeV, which only allows to resolve thresholds for up to $\sqrt{s}/2$.
\label{fig:3point}}
\end{figure*}

Let us start by discussing the main features of di-Higgs production. The effect of the new scalars is mainly encoded in the modification of the three-point function.
A comparison to the Standard Model value is shown in Fig.~\ref{fig:3point}.\footnote{Although the quantity shown in Fig.~\ref{fig:3point} is not an observable, it is instructive to understand where corrections can be expected in physical quantities derived from $\Gamma$.}
For relatively low masses of the new scalar this exhibits a considerable momentum dependence as well as featuring a real and an imaginary part.
Both properties distinguish the full calculation from estimates based on a modified three-Higgs boson coupling obtained from the (explicitly momentum-independent) Coleman-Weinberg effective potential~\cite{Noble:2007kk,Curtin:2014jma}.
From Fig.~\ref{fig:3point} it also becomes clear that we quickly probe the decoupling limit for larger values of $m_S$ (consistent with the expectation of \cite{Appelquist:1974tg}).

\begin{figure}[!t]
\centering
\includegraphics[width=0.4\textwidth]{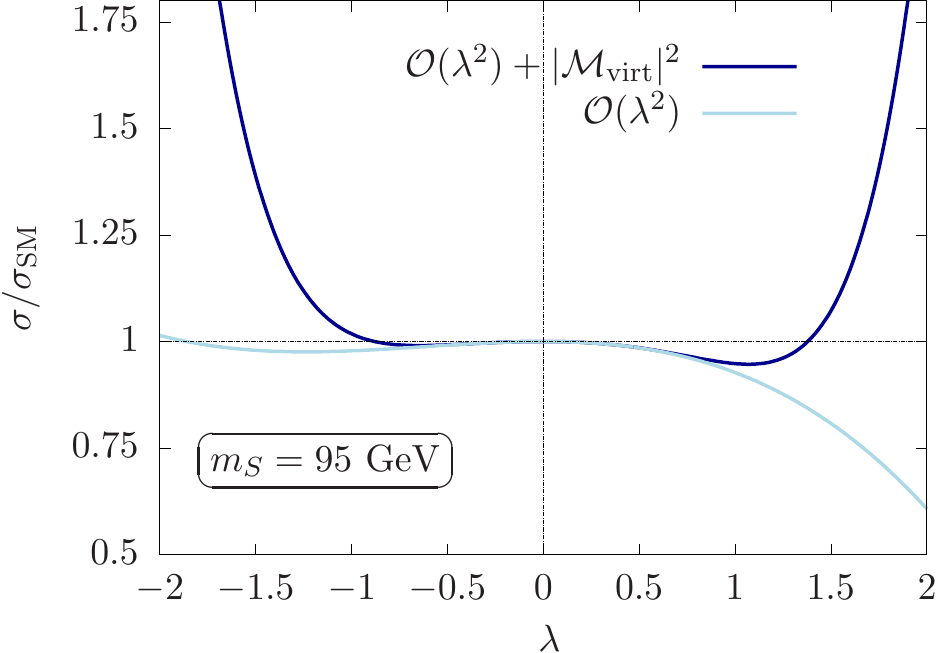}
\caption{Comparison of the cross section as a function of $\lambda$ for different approximations, $m_S=95~\text{GeV}$ and $\sqrt{s}=100~\text{TeV}$, for details see text.\label{fig:mass}}
\end{figure}

The price to pay for using a loop process is that it is higher order in the portal coupling $\lambda$. In practice we find that sensitivity is limited to $\lambda\sim 1$
where these effects are accordingly non-negligible. At the same time this raises the question of even higher order corrections.
As already mentioned  in the previous section, we can use the ${\cal{O}}(\lambda^4)$ term $|{\cal{M}}_{\text{virt}}|^2$ to obtain some estimate of higher order corrections beyond the  ${\cal{O}}(\lambda^3)$ we have fully included. A comparison is shown in Fig.~\ref{fig:mass} giving us reasonable confidence in the calculations for couplings up to $\lambda\sim 1$, especially for light scalar masses.

\begin{figure}[!t]
\centering
\includegraphics[width=0.4\textwidth]{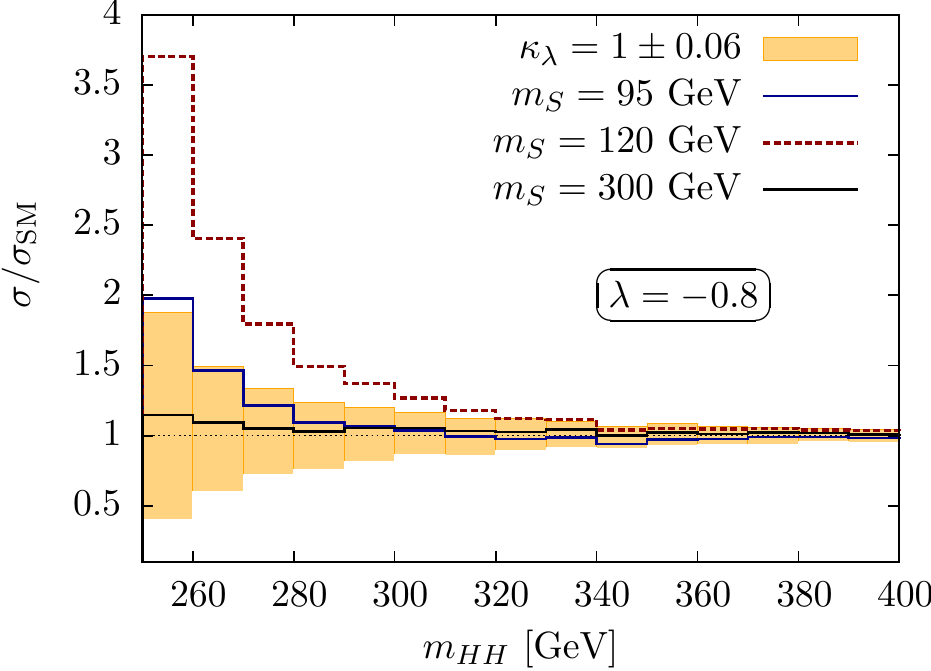}
\caption{
Invariant di-Higgs mass spectra relative to the SM and 6\% self-coupling extraction as described in~\cite{Contino:2016spe}.
\label{fig:hhspecb}}
\end{figure}

Let us now turn to the actual evaluation of the sensitivity.
Most projections for the precision of di-Higgs rate measurements are quantified as sensitivities to changes in the Higgs boson self-coupling (cf., e.g.,~\cite{Contino:2016spe,CMS:2018ccd}).
To make use of these, we compare the impact of virtual portal scalars against that of a (momentum independent) change in the self-coupling as shown in Fig.~\ref{fig:hhspecb}.
If the binned distribution deviates by more than the band indicated by the self-coupling projection in the sense of a binned $\chi^2$ test, we consider a particular $(m_S,\lambda)$ point to be excludable.

%

\begin{figure}[!t]
\centering
\includegraphics[width=0.44\textwidth]{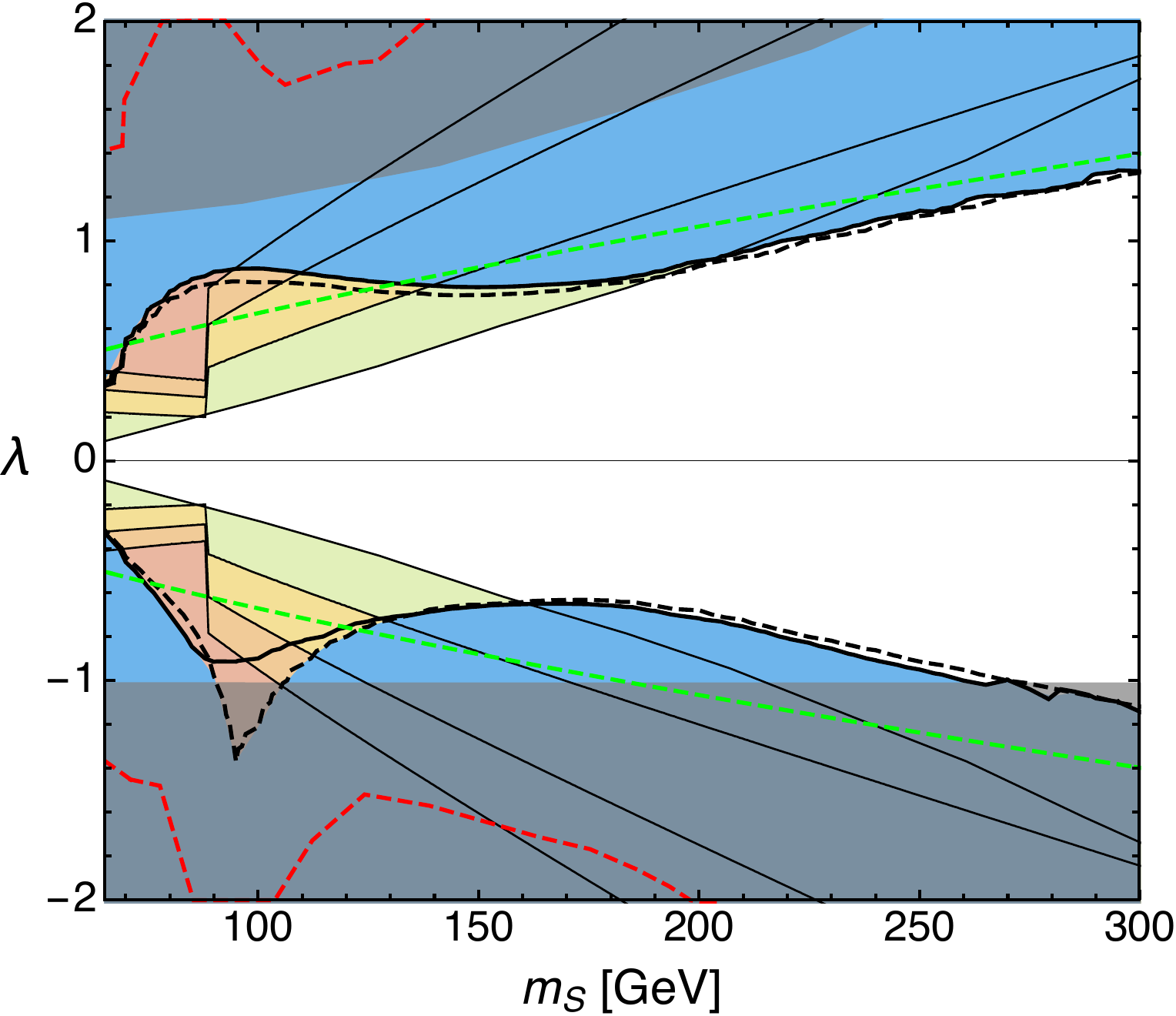}
\caption{Sensitivity projections for the di-Higgs boson production as well as other approaches. 
LHC Di-Higgs results are shown as the red dashed line. For FCC-hh (100 TeV, 30/ab) we show as the solid (dashed line) the sensitivities obtainable from di-Higgs boson production based on  a calculation without (with) terms $\sim |{\cal{M}}_{\text{virt}}|^2$. The blue region shows the intersection of these two calculations and is this sense conservative. The green dashed line shows the sensitivity expected from a simple effective potential approximation for the self-coupling~\cite{Noble:2007kk,Curtin:2014jma}. The red, orange and yellow regions corresponds to a 1.7\%, 1.06 \%  and 0.5\% measurement~\cite{deBlas:2019rxi} of the cross section of vector boson associated Higgs boson production at a future ILC-250 (International Linear Collider at 250 GeV), CLIC-380 (Compact Linear Collider at 380 GeV) and FCC-ee (electron--positron option of the FCC at 240 GeV), respectively. Finally the light green region indicates the best sensitivity curve from missing energy searches according to~\cite{Craig:2014lda}. Regions, obtained in~\cite{Curtin:2014jma}, where the electroweak vacuum is potentially endangered  are shown in grey.
\label{fig:results}}
\end{figure}

We consider both the sensitivity at LHC but also a future FCC.
The implicit momentum dependence of $pp\to HH$ has been used to set constraints on the Higgs boson self-coupling by exploiting the destructive interference between the triangle and box contributions of Fig.~\ref{fig:lotops}. Given the relatively small cross section of $HH$ production at the LHC of about $32$~fb~\cite{Dawson:1998py,Frederix:2014hta,deFlorian:2015moa,deFlorian:2016uhr,Borowka:2016ehy,Borowka:2016ypz,Heinrich:2017kxx,Grazzini:2018bsd,deFlorian:2018tah,Baglio:2018lrj}, the expected precision of the self-coupling extraction is going to be limited. A recent projection by CMS~\cite{CMS:2018ccd} suggests that a sensitivity to $\lambda^{95\%\,\text{CL}}_{\text{SM}}/\lambda_{\text{SM}}= [-0.18, 3.6] $ can be achieved, which corresponds to a gluon fusion cross section extraction of ${\cal{O}}(15\%)$ when assuming SM dynamics.
The obtainable sensitivity is shown as the red dashed line in Fig.~\ref{fig:results}. As we can see, detectable effects typically require couplings $\lambda$ significantly larger than $1$, where our calculations are not fully trustworthy. To be conservative we perform the calculation with and without the squared virtual corrections and only show whatever sensitivity is weaker. However, it should be kept in mind that this still includes only part of the higher order corrections and therefore is only an estimate.

Di-Higgs boson production is one of the key motivators for pushing the high-energy frontier beyond the high-luminosity and high-energy LHC options. As shown in Ref.~\cite{Contino:2016spe} (see also \cite{Yao:2013ika,Barr:2014sga,He:2015spf,Papaefstathiou:2015iba,Adhikary:2017jtu,Banerjee:2018yxy,Banerjee:2019jys}) a coupling extraction of $\lambda_\text{SM}$ at the $\lesssim 6\%$~level could be attainable at an FCC-hh with 100 TeV collisions and a 30/ab dataset. This is a direct reflection of a much larger di-Higgs inclusive cross section of around 1~pb~\cite{Grazzini:2018bsd}.
On the basis of this extrapolation, a much better sensitivity to the portal coupling can be achieved. This is shown as the solid black line and the blue region in Fig.~\ref{fig:results} which now penetrates into the region $\lambda\lesssim 1$. To estimate the sensitivity to higher order corrections we show the dashed line that includes the squared virtual corrections. 

The impact of the full calculation can be appreciated by comparison with the green dashed line which is obtained from estimating the change in the Higgs' self-coupling by including the effects of the portal scalars in the Coleman-Weinberg effective potential as in~\cite{Noble:2007kk,Curtin:2014jma}.\footnote{The Coleman-Weinberg effective potential~\cite{Coleman:1973jx} is given by $V(H)=-(\mu^2/2)H^2+(\lambda_{H}/4)H^4+(1/(64\pi^2))M^4(\text{const.}+\log(M^2))$, where in our case $M^2=m^{2}_{S}+\lambda(H^2-v^2)$. As in~\cite{Noble:2007kk} we have fixed $\mu^2$ and $\lambda_{H}$ by implementing the condition that $v=246\,{\rm GeV}$ and $m^{2}_{H}=(125\,{\rm GeV})^2$. The results agree well with those of~\cite{Curtin:2014jma}.} 
The alert reader might realise that, for larger values of $m_S$, the full computation results in a systematically higher sensitivity than the effective potential calculation, where we could expect the Coleman-Weinberg approximation to be a good one according to Fig.~\ref{fig:3point}. 
The reason for this difference is that the expected precision as given by the $\lambda_H/\lambda_H^{\text{SM}}\sim {\cal{O}}(6\%)$ interpretation pushes us into a regime where weak corrections become relevant. These are not fully reflected by ad-hoc rescalings of the Higgs boson self-coupling. Tree-level modifications $\lambda_H/\lambda_H^{\text{SM}}$ are visible through threshold effects~\cite{Glover:1987nx,Baur:2002rb,Baur:2002qd,Baur:2003gpa} (see also \cite{Barr:2014sga}). Effects of this type need to be contrasted with coherent Higgs boson coupling changes~\cite{Englert:2013tya,Craig:2013xia} that drive the destructive interference between the diagrams in Fig.~\ref{fig:lotops}, in particular they affect the box diagrams. As the latter contributions are relevant even in the high $m_{HH}$ region where FCC-hh has significant sensitivity, there is an additional source of deviation compared to $\lambda_H/\lambda_H^{\text{SM}}$ alone. The eventual sensitivity yield will obviously depend on the details of the machine itself as well as the status of SM precision calculations at the time. That said, it is clear that sufficiently large statistics could enable us to go beyond just finding a deviation from the SM and fingerprint the origin of the changes in the invariant di-Higgs boson mass spectrum (see also the recent~\cite{Capozi:2019xsi}). As can be seen from Fig.~\ref{fig:hhspecb}, the dependence on the invariant mass for the portal scalar is quite different from a simple change in the Higgs boson self-coupling allowing to get information on the new physics giving rise to the deviation from the SM.

\bigskip

The sensitivity of di-Higgs boson production has to be appraised in the context of other approaches that have been suggested to constrain the model of Eq.~\eqref{eq:model}. 
In the following we concentrate on two main methods: the change in the cross section of vector boson associated Higgs boson production (Higgs-strahlung)~\cite{Englert:2013tya,Craig:2013xia,Craig:2014una} (see also \cite{Curtin:2014jma}) and processes where the new scalars $S$ are produced via an off-shell Higgs, typically leading to missing energy~\cite{Craig:2014lda,Arcadi:2019lka}.

It is known that the associated weak corrections will modify the single Higgs boson production phenomenology~\hbox{\cite{Englert:2013tya,Craig:2013xia,Craig:2014una}} leading to measurable deviations in particular at future precision machines such as a future lepton collider, FCC-ee, in $Z$ boson-associated Higgs production. These constraints do not depend on the energy momentum transfer, also because the measurement will be focussed on a very narrow energy range of around 240 GeV~\cite{Peskin:2012we,Gomez-Ceballos:2013zzn,Klute:2013cx,Abada:2019zxq}, where $Z$ boson-associate Higgs production is maximised. The energy spectrum of the incident electrons is typically sharp~(see e.g.~\cite{Boogert:2002jr}) for regions where $e^+e^-\to ZH$ production is relevant.
The yellow region in Fig.~\ref{fig:results} shows the sensitivity obtainable with an $0.5\%$~\cite{Klute:2013cx} precision measurement of the cross section. We have performed similar analysis for the ILC-250 (1.06\%)~\cite{deBlas:2019rxi} and CLIC-380 (1.7\%)~\cite{deBlas:2019rxi} that are shown in orange and red, respectively.

At hadron colliders, the Higgs portal interaction leads to pair production of the new scalar $S$ via an off-shell Higgs boson, giving rise to 
a missing energy signature~\cite{Craig:2014lda,Arcadi:2019lka}.  Such analyses are difficult as no resonance structure is available to control backgrounds. In parallel, the cross section has a steep drop-off as a function of the Higgs bosons' virtuality due to the propagator suppression. However, we can expect considerable sensitivity, in particular when we turn to 100 TeV collisions with large statistics 30/ab~\cite{Craig:2014lda}, which is given by the light green region in Fig.~\ref{fig:results}. Another possibility at hadron colliders is off-shell Higgs boson production in $Z$ pair final states. The non-decoupling of the Higgs contribution due to unitarity arguments~\cite{Kauer:2012hd} has been used to set constraints on the Higgs boson width~\cite{Caola:2013yja} under certain assumptions~\cite{Englert:2014aca,Englert:2014ffa}. It has been shown that a fair part of corrections cancel in this channel~\cite{Englert:2019zmt} and an explicit calculation shows that only sub-percent modifications of the Higgs signal in $ZZ$ can be expected in the region selected by the $HH$ production projection. This small modification needs to be contrasted with a dominant $gg\to ZZ$ continuum which makes $HH$ production the more relevant channel in the light of the associated 3-point modification that is magnified by the interference effects in the $HH$ threshold region.

Finally we also remark that a viable electroweak symmetry breaking vacuum is endangered if the portal coupling is too large~\cite{Curtin:2014jma}. We indicate the region where non-perturbatively large self-couplings of the scalar $S$ are needed to avoid this fate according to~\cite{Curtin:2014jma} as the grey region. This shows that the LHC's sensitivity in $HH$ production is not large enough to test a viable region of the model's parameter space in this channel.

\section{Summary and Conclusions}
Higgs physics remains an exciting avenue to explore the potential presence of new interactions beyond the Standard Model. In particular, given the 
gauge-singlet nature of the $|\Phi|^2$ operator, a fully renormalisable scalar (Higgs portal) extension $\sim |\Phi|^2 S^2$ of the SM is a motivated possibility. While such models can be efficiently constrained when the additional scalar obtains a vacuum expectation value~\cite{Englert:2011yb,Bertolini:2012gu,Bernaciak:2014pna,Robens:2015gla} or is light enough for the decay $H\to SS$ to be open (cf., e.g.,~\cite{Arcadi:2019lka,Cepeda:2019klc} for a recent discussion), once $m_S>m_H/2$ the sensitivity becomes limited directly as a consequence of the (weak) Higgs-related production. At hadron colliders, off-shell Higgs boson measurements can provide direct sensitivity~\cite{Craig:2014lda} that is limited ultimately by the reduction of the cross section for masses too far away from the Higgs resonance. In these circumstances, i.e. when the probed centre-of-mass energy is high enough, these scalars can manifest themselves as virtual contributions through (but not limited to) absorptive parts of the amplitude. This motivates the precision study of double Higgs boson final states (see also~\cite{Voigt:2017vfz}) as an indirect probe which is expected to become a sensitive probe of electroweak physics at a future high-energy proton collider. In this work we have shown that the expected precision of the Higgs self-coupling extraction at a 100 TeV FCC-hh~\cite{Contino:2016spe} indeed shows competitive (yet model-dependent) sensitivity to this scenario, with complementarity to precision studies of Higgs-strahlung processes at, e.g., lepton colliders \cite{Craig:2013xia,Craig:2014una}. This shows that the energy coverage and the large anticipated data set at such a machine can provide a competitive electroweak precision physics instrument (see also \cite{Franceschini:2017xkh,Banerjee:2018bio}).

\bigskip 

\noindent {\textbf{Acknowledgements}} --- 
We thank Ian Lewis, Shi-Ping He, Tilman Plehn, Michael Spannowsky and Susanne Westhoff for helpful discussions and comments on the manuscript.
C.E. would like to thank the Institute for Theoretical Physics Heidelberg for their hospitality during an early stage of this work. 
C.E. is supported by the UK Science and Technology Facilities Council (STFC) under grant ST/P000746/1. 
J.J. would like to thank the IPPP for hospitality and gratefully acknowledges support under the DIVA fellowship program of the IPPP.

\bibliography{paper.bbl} 

\end{document}